\documentclass[a4paper]{article}
\usepackage{amsfonts}
\usepackage{amssymb}
\usepackage[margin=2cm]{geometry}
\usepackage[dvips]{graphicx}

\usepackage[T1]{fontenc}
\usepackage[latin2]{inputenc}

\usepackage{amsmath}
\usepackage{bbm}

\begin{document}


\title{False qubits?: Polarization of light and Josephson junction}

\author{Robert Alicki \\ 
  {\small
Institute of Theoretical Physics and Astrophysics, University
of Gda\'nsk,  Wita Stwosza 57, PL 80-952 Gda\'nsk, Poland}\\
}

\date{\today}
\maketitle

\begin{abstract}
We compare two physical systems: polarization degrees of freedom of a macroscopic light beam and the Josephson junction (JJ) in the "charge qubit regime". The first system obviously cannot carry genuine quantum information and we show that the maximal entanglement which could be encoded into polarization of two light beams scales like 1/(photon number). Two theories of JJ, one leading to the picture of "JJ-qubit" and the other based on the mean-field approach are discussed. The later, which seems to be more appropriate, implies that the JJ system is, essentially,  mathematically equivalent to the polarization of  a light beam with the number of photons replaced by the number of Cooper pairs. The existing experiments consistent with the "JJ-qubit" picture and the theoretical arguments supporting, on the contrary, the classical model are briefly discussed. The Franck-Hertz-type experiment is suggested as an ultimate test of the JJ nature. 

\end{abstract}

In the year 1852 Stokes proposed to describe the state of light beam polarization by a set of four  real parameters
$I,M, C, S $ satisfying the condition $I\geq \sqrt{M^2+C^2+S^2}$. One can associate with those parameters a $2\times 2$ "density
matrix" normalized to the intensity of the beam $I$ as follows
\begin{equation}
{\hat{\Omega}} =  \frac{1}{2}(I{\hat\sigma}_0+ M{\hat\sigma}_x + C{\hat\sigma}_y+ S{\hat\sigma}_z)\ .
\label{stokes}
\end{equation} 
Then, within the validity of the linear optics, any action of the optical device can be described by a completely positive and generally trace decreasing map (see Mueller and Jones calculus \cite{pol}) which transforms the input polarization state into the output one
\begin{equation}
{\hat\Omega}_{out} = \Lambda ({\hat\Omega}_{in}) \ .
\label{mueller}
\end{equation} 
The Stokes parameters can be measured by applying the procedure analogical to the "tomography of a qubit state".
The formal analogy with the description of  states of the 2-level system suggests the following question: \emph{Can polarization state of a light beam encode a qubit?}\\
To answer this question we consider the quantum origin of the "density matrix" $\Omega$.
It is in fact a correlation matrix which can be written in terms of the quantum average
\begin{equation}
{\Omega}_{\mu\nu} = {\rm Tr}\bigl({\hat\rho}\, {\hat a}^{\dagger}_{\nu} {\hat a}_{\mu}\bigr) \ .
\label{corr}
\end{equation} 
Here ${\hat\rho}$ is a density matrix describing quantum electromagnetic field of the beam and  ${\hat a}^{\dagger}_{\mu},  {\hat a}_{\mu}$
are creation and anihilation operators for the mode $\mu$. In particular for a monochromatic beam with a fixed wave vector ${\bf k}$ the indices $\mu,\nu = 1,2$ correspond to different polarization basis and such ${\hat\Omega}$
is a positively defined matrix normalized to the averaged number of photons in the beam and equivalent to the Stokes matrix (\ref{stokes}). The correlation matrix (\ref{corr}) allows to compute  mean values of the additive observables described by the operators of the form  $K= \sum k_{\mu\nu}a^{\dagger}_{\nu} {\hat a}_{\mu}$.

If our beam consists always of a single photon, then it corresponds to a qubit and the entangled states of two photons can be produced \cite{Tit}. Obviously, this is the only case when  polarization describes a true qubit. To see how those quantum properties  vanish with the increasing number of photons we consider two beams with the associated creation and annihilation operators ${\hat a}^{\dagger}_{\mu},  {\hat a}_{\mu}, {\hat b}^{\dagger}_{\mu'},  {\hat b}_{\mu'}$, respectively. The correlation matrix which could now correspond to "2- qubit density matrix" is given by
\begin{equation}
\Gamma_{\mu\mu',\nu\nu'} = {\rm Tr}\bigl({\hat\rho}\, {\hat a}^{\dagger}_{\nu} {\hat a}_{\mu}{\hat b}^{\dagger}_{\nu'}  {\hat b}_{\mu'}\bigr) 
\label{corr2}
\end{equation}
and is normalized to the averaged product of photon numbers  $\overline{n_a n_b}$. Denoting the normalized versions of
(\ref{corr2}) by $\tilde{\Gamma}$ we can compute the upper bound for the \emph{amount of entanglement} expressed in terms of  \emph{negativity} \cite{Pl} $N(\hat{\sigma}) =\frac{1}{2}( {\rm Tr} (|{\hat\sigma}^{\Gamma}|)-1)$ where ${\hat\sigma}^{\Gamma}$ is a partially transposed 2-qubit density matrix. We can write now
\begin{equation}
\tilde{\Gamma}^{\Gamma}_{\mu\mu',\nu\nu'} = \frac{1}{\overline{n_a n_b}}{\rm Tr}\bigl({\hat\rho}\, {\hat a}^{\dagger}_{\nu} {\hat a}_{\mu}{\hat b}^{\dagger}_{\mu'}  {\hat b}_{\nu'}\bigr)
= \frac{1}{\overline{n_a n_b}}{\rm Tr}\bigl({\hat\rho}\,({\hat a}_{\nu}{\hat b}^{\dagger}_{\nu'})^{\dagger} ({\hat a}_{\mu}{\hat b}^{\dagger}_{\mu'})\bigr)
- \frac{1}{\overline{n_a n_b}}{\rm Tr}\bigl({\hat\rho}\, {\hat a}^{\dagger}_{\nu} {\hat a}_{\mu}\bigr)\delta_{\mu'\nu'}\ .
\label{pt}
\end{equation}
As the first matrix on the RHS of (\ref{pt}) is again a correlation matrix and hence positively defined we have
\begin{equation}
{\rm Tr}\bigl(|\tilde{\Gamma}^{\Gamma}|\bigr)\leq  1 +\frac{4{\bar n}_a}{\overline{n_a n_b}}
\label{pt1}
\end{equation}
what yields the upper bound on the value of negativity 
\begin{equation}
N(\tilde{\Gamma}) \leq \frac{2{\bar n}_a}{\overline{n_a n_b}}\simeq \frac{2}{{\bar n}_b}\ .
\label{neg}
\end{equation}
As the estimation is symmetric with respect to both beams the upper bound on negativity is inversely proportional to the largest averaged photon number. One should remember that for our case of $2\times 2$ density matrices the condition
$N({\hat\sigma})> 0$ is a necessary and sufficient one for the entanglement \cite{H3}.
\par
The results of the above analysis seem to be rather obvious, as nobody expects to realize a qubit using polarization of a macroscopic light beam. The next example is not obvious at all. The implementation of quantum information processing based on the so-called \emph{superconducting qubits} is considered to be quite promissing \cite{D},\cite{W}. Nevertheless, we shall argue that the encoding of quantum information in superconducting qubits based on Josephson junctions meets the same fundamental restrictions as in the case of macroscopic light polarization. 
\par
We consider for simplicity the version of JJ called "charge qubit" but the general results
remain true for other cases also. A system of two superconducting electrodes "1" and "2" separated by a thin layer of an insulator  allows for tunneling of Cooper pairs which are treated as a bosonic gas under the Bose-Einstein condensation conditions. The electrode "1" is assumed to be small enough to make Coulomb interaction between Cooper pairs important. We describe the system by  creation and annihilation operators ${\hat a}^{\dagger}_{\mu},  {\hat a}_{\mu}, \mu=1,2$ corresponding to the ground states of a Cooper pair in the corresponding electrodes. The condition of the Bose-Einstein condensation for Cooper pairs is necessary to achieve macroscopic occupation of the lowest energy levels in both electrodes. This phase-transition makes a system of massive particles similar to a system of photons which, as massles, can always macroscopically occupy a single mode. The later property follows formally from the relation:
BEC-temperature $\sim$ 1/boson's mass. The Hamiltonian describing the dynamics of the condensate is the following
\begin{equation}
{\hat H} = {\hat H}_C  + \frac{\lambda}{2} ({\hat a}_1 {\hat a}_2^{\dagger} + {\hat a}_1^{\dagger}{\hat a}_2)
\label{JJ}
\end{equation}
where ${\hat H}_C$ is responsible for the Coulomb interaction and the term proportional to $\lambda$ describes the tunneling
of Cooper pairs. In order to model the Coulomb interaction we "quantize" the classical expression for the energy of a capacitor 
\begin{equation}
E_{cap} = \frac{Q^2}{2C}
\label{cap}
\end{equation}
where $Q$ is an excess charge and $C$ is a capacity of the electrode "1".
The first method is based on the formal substitution   $ Q\mapsto -2e [{\hat a}_1^{\dagger}{\hat a}_1 - {\bar n}_1]$ where  ${\bar n}_1$ is an averaged background number of Cooper pairs at the electrode "1" and leads to the Hamiltonian
\begin{equation}
{\hat H}_C = E_C [{\hat a}_1^{\dagger}{\hat a}_1 - {\bar n}_1]^2
\label{bh}
\end{equation}
with $ E_C= 2e^2/C$. Using the relation $|n-{\bar n}_1 | << {\bar n}_1 << N$, satisfied in the relevant "charge qubit regime",  we can approximate (\ref{bh}) by the Bose-Hubbard Hamiltonian which leads directly to the popular model of a "macroscopic quantum system" \cite{S},\cite{RA}.  Such a quantum device should exhibit phase fluctuations of the order ${\cal O}(1/\sqrt{\Delta n})$ and charge fluctuations of the order   ${\cal{O}}(\sqrt{\Delta n})$ where $\Delta n$ is a typical value of the excess number of pairs $(Q = -2e\Delta n)$.

There is another choice - the mean-field (Hartree-type) nonlinear (state-dependent) Hamiltonian
\begin{equation}
{\hat H}_C = E_C [\langle {\hat a}_1^{\dagger}{\hat a}_1\rangle - {\bar n}_1][{\hat a}_1^{\dagger}{\hat a}_1 - {\bar n}_1]
\label{m-f}
\end{equation}
where $\langle \cdot\rangle$ denotes the quantum average with respect to the actual state.
There exist a number of physical and mathematical arguments supporting the mean-field form (\ref{m-f}). First of all the mean field Hamiltonian (\ref{m-f}) provides a more realistic picture of a quasi-particle  charge feeling the averaged potential produced by all other quasi-particle charges occupying the electrode "1". As their number is large (${\bar n}_1\sim 10^9$), the structure of quasi-particle is strongly delocalized and the Coulomb interaction is a long-range one, the mean-field approximation seems to be more appropriate. It is also important that due to superconducting phase transition the Cooper pairs occupy a single quantum state. It follows that the density of the condensate is essentially a classical variable with quantum fluctuations of the order $1/\sqrt{{\bar n}_1}$.  To deal with such situations  all standard theories use various types of mean-field methods. In the Bogoliubov approach to superconductivity or superfluidity \cite{Bog} the interaction Hamiltonian terms quatric in field operators $(\psi^{\dagger}\psi)(\psi^{\dagger}\psi)$ are replaced by bilinear expressions with state -dependent coefficients  $\langle\psi^{\dagger}\psi\rangle\psi^{\dagger}\psi$. Similar self-consistent approximations are fundamental for the derivations of the Landau-Ginzburg and the Gross-Pitaevski equations \cite{GP}, \cite{E}.

The Hartree-type Hamiltonian (\ref{m-f}) preserves the product structure of the states  of Cooper pairs distributed coherently among  two electrodes (we omit the irrelevant overall phase factor)
\begin{equation}
|n,\phi\rangle =  \frac{1}{\sqrt{N!}}\Bigl[\sqrt{\frac{n}{N}}e^{i\phi}{\hat a}_1^{\dagger} + \sqrt{\frac{N-n}{N}}{\hat a}_2^{\dagger}\Bigr]^N|vac\rangle
\label{prod}
\end{equation}
where $N$ is a total number of Cooper pairs,  $n$ is their average number at the electrode "1" and $\phi$ is the relative phase. Using again the relation $|n-{\bar n}_1 | << {\bar n}_1 << N$ to simplify the formulas we obtain
the following  evolution equations for $\phi$ (pendulum equation)\cite{pen}  
\begin{equation}
{\ddot{\phi}} = - \omega^2 \sin \phi 
\label{pend}
\end{equation}
where $E_J = \lambda\sqrt{{\bar n}_1(N-{\bar n}_1)}$, $\omega^2
= 2 E_C E_J/\hbar^2$ and the additional relation $\dot{\phi}= (E_C/\hbar)[n-{\bar n}_1]$ holds.

The product structure of the state (\ref{prod}) implies, for large $N$, normal fluctuations of the phase $\phi$ and the quasi-particle number $n$ of the order $1/\sqrt{{\bar n}_1}$ and $\sqrt{{\bar n}_1}$, respectively. This agrees with the standard picture of the phase $\phi$ being an order parameter associated with the superconductivity phase transition which becomes a classical observable in the limit $n\to\infty$. Such an observable should display normal fluctuations except for the case of the critical temperature. Quadratic in creation and annihilation operators structure of the mean-field Hamiltonian (\ref{m-f}) implies also the existence of a closed but nonlinear evolution equation for the correlation matrix ${\Omega}_{\mu\nu} = {\rm Tr}\bigl({\hat\rho}\, {\hat a}^{\dagger}_{\nu} {\hat a}_{\mu}\bigr)$. This simple, "single-particle" structure of the evolution is preserved even for the model of JJ device interacting with an environment, if the main source of dissipation is escape and return of quasi-particles to the condensate phase. Therefore a nonlinear version of the dynamical map (\ref{mueller}) makes sense for JJ also. In fact the analogy to a light beam is even closer. If we take into account photon-photon scattering predicted by the quantum electrodynamics the map (\ref{mueller}) must be also nonlinear. The fact that the measurable quantities of JJ are completely determined by the correlation matrix ${\Omega}_{\mu\nu}$,  similarly to the polarization of a light beam, implies that the estimation of the amount of entanglement which can be encoded into a pair of two light beams (\ref{neg}) can be applied for two JJ devices as well.

The arguments of above supporting the picture of JJ as an essentially classical system seem to contradict the existing experiments \cite{NBS}. In particular those experiments show coherent oscillations of a charge, spectral evidence of the coupling between two JJ's and, finally the "entanglement of the superconducting qubits via state tomography". One should, however, notice that the experimental results are merely \emph{consistent} with the mathematical model based on the quantum picture of JJ. It does not mean that they cannot be also consistent with the classical JJ model and, indeed, the numerical analysis of the classical microwave-driven JJ presented in \cite{Gro} show that this is the case. Unfortunately, it is not easy to design an experiment which could ultimately reject one of the models. The natural candidate - model-independent test of Bell inequalities- is rather nonconclusive because of the strong presence of the "locality loophole" in the case of coupled JJ devices \cite{loop}. Perhaps, the old idea of the Franck-Hertz experiment
could be implemented here. Assume, we can couple a single JJ device, working in the regime corresponding to a "few-level quantum system", to another system with a continuous energy spectrum. If this energy would be absorbed  by the JJ device in quantized  portions $\hbar\omega$ then the classical model should be dismissed.

\par
\emph{Acknowledgements.}  The author is grateful Marco Piani, Frank Wilhelm and Robert Raussendorf for discusions. Financial support 
by the Polish Ministry of Science and Information
Technology - grant PBZ-MIN-008/P03/2003 and by the European Union through the Integrated Project SCALA is acknowledged.


\begin{thebibliography}{99}
\bibitem{loop} This fact was pointed out to the author by Daniel Gottesman.
\end{thebibliography}
\end{document}